\documentclass[twocolumn,superscriptaddress,aps,prb]{revtex4}
\usepackage{amsmath,amssymb, amsthm}
\usepackage{graphicx}
\usepackage{dcolumn}
\usepackage{color}
\usepackage{bm}
\usepackage{ulem}

\begin{document}

\title{Many-Body Localization from Dynamical Gauge Fields}
\author{Zhiyuan Yao}
\affiliation{Institute for Advanced Study, Tsinghua University, Beijing, 100084, China}

\author {Chang Liu}
\affiliation{Institute for Advanced Study, Tsinghua University, Beijing, 100084, China}

\author {Pengfei Zhang}
\affiliation{Institute for Advanced Study, Tsinghua University, Beijing, 100084, China}
\affiliation{Institute for Quantum Information and Matter, California Institute of Technology, Pasadena, California 91125, USA}
\affiliation{Walter Burke Institute for Theoretical Physics, California Institute of Technology, Pasadena, California 91125, USA}

\author {Hui Zhai}
\email{hzhai@tsinghua.edu.cn}
\affiliation{Institute for Advanced Study, Tsinghua University, Beijing, 100084, China}
\date{\today}

\begin{abstract}
    A recent experiment [Nature Physics \textbf{10}, 1 (2019)] has realized a dynamical gauge system with $\mathbb{Z}_2$ gauge symmetry in a double-well potential. In this work we propose a method to generalize this model from a single double well to a one-dimensional chain. We show that although there is no disordered potential in the original model, the phenomenon of many-body localization can occur. The key ingredient is that different symmetry sectors with different local gauge charges play the role of different disorder configurations, which becomes clear after exactly mapping our model to a transverse Ising model in a random longitudinal field. We show that both the ergodic regime and the many-body localized regime exist in this model from four different metrics, which include level statistics, volume law versus area law of entanglement entropy of eigenstates, quench dynamics of entanglement entropy and physical observables.
\end{abstract}

\maketitle

\section{Introduction}

In the past decade, one of the main research topics in cold atom physics is simulating synthetic gauge fields \cite{spielman_physics_today,spielman_review,dalibard_RMP}. The major tool is either using Raman transitions or  shaking optical lattices periodically. It first started with simulating a constant abelian gauge field \cite{spielman_constant_gauge,sengstock_gauge}, which can be gauged out and does not have physical effects.  Physical effects of gauge fields can be produced either by introducing a spatial or temporal dependence \cite{spielman_magnetic_field,spielman_electric_field} or by generalizing gauge fields from abelian ones to non-abelian ones \cite{spielman_SOC,Zhang_SOC_fermion,MIT_SOC_fermion}. By introducing spatial or temporal dependence, this protocol realizes synthetic magnetic fields \cite{spielman_magnetic_field} or synthetic electric fields \cite{spielman_electric_field}, which manifest themselves as vortices in superfluids \cite{spielman_magnetic_field} and the Hall effect \cite{spielman_Hall}. By generalizing to the non-abelian case, spin-orbit couplings can be induced \cite{spielman_SOC_review,Zhai_SOC_review1,Zhai_SOC_review2}, which give rise to rich physics such as the stripe superfluids for Bose condensates \cite{Zhai_stripe, Ho_stripe, stringari_stripe, USTC_stripe,Ketterle_stripe} and topological bands for Fermi gases \cite{Cooper_review}. Nevertheless, the primary focuses in these studies are the properties of matter fields since the gauge fields are fixed by external classical sources, such as laser fields and magnetic fields, and do not receive backaction from the matter fields.

In the second wave of research along this line, the gauge fields acquire their own dynamics. For example, recently density-dependent gauge fields have been created using lattice shaking, either by tuning the shaking frequency resonant with the interaction strength \cite{ETH_density_dependent_gauge} or by periodically driving both the optical lattice and the interaction at the same frequency \cite{Cheng_density_dependent_gauge}. As we know, density of a many-body system is a dynamical variable that can fluctuate spatially and temporally. Therefore, such density-dependent gauge fields are dynamical. Nevertheless, this does not mean they are ``dynamical gauge fields" as we understand in high energy physics,  where local gauge symmetries are the essential ingredients. In other words, to be eligible for dynamical gauge fields, not only the gauge fields need to have their own dynamics, but also the dynamical terms have to be invariant under local gauge transformations. One familiar example is the Maxwell theory in which the dynamical term $-\frac{1}{4} \mathcal{F}_{\mu\nu}\mathcal{F}^{\mu\nu}$, where $\mathcal{F}^{\mu\nu}$ is the electromagnetic field tensor, is invariant under local $U(1)$ gauge transformations. 

Since local gauge symmetries are missing in the above-mentioned density-dependent gauge field systems, strictly speaking, dynamical gauge fields with local gauge symmetries have not been synthesized in cold atom systems. However, significant progresses have been made in recent experiments, and a zero-dimensional two-site $\mathbb{Z}_2$ version of dynamical gauge fields has been created \cite{Bloch_Z2_gauge}. This model possesses a $\mathbb{Z}_2$ gauge symmetry, although there is no distinction between ``local" and ``global" in the two-site case. In the experiment reported in Ref. \cite{Bloch_Z2_gauge}, two ${}^{87}$Rb atoms, one in hyperfine state $\vert F=1, m_{F}=1 \rangle$ ($f$-atom) and one in hyperfine state $\vert F=1, m_{F}=-1 \rangle$  ($a$-atom), are confined in a double-well potential. The creation and annihilation operators of $f$-atom ($a$-atom) are denoted by $\hat{f}^\dag_i$ ($\hat{a}^\dag_i$) and $\hat{f}_i$ ($\hat{a}_i$), with $i=1,2$ for the two sites. Since the total number of $f$-atom in the double well is conserved and fixed as one, we can introduce a spin-$1/2$ operator $\hat{\bm{\tau}}_{i}$ for this double well defined as follows ($\hbar = 1$)
\begin{align}
    \hat{\tau}^x=\frac{1}{2}\left(\hat{f}^\dag_1\hat{f}_2+\hat{f}^\dag_2\hat{f}_1\right), \label{tau_x} \\
    \hat{\tau}^y=\frac{i}{2}\left(\hat{f}^\dag_2\hat{f}_1-\hat{f}^\dag_1\hat{f}_2\right), \label{tau_y} \\
    \hat{\tau}^z=\frac{1}{2}\left(\hat{f}^\dag_1\hat{f}_1-\hat{f}^\dag_2\hat{f}_2\right).\label{tau_z}
\end{align}
By shaking the optical lattice at some fine-tuned frequency, one can reach the following effective Hamiltonian to describe the coupling between the $a$-atom and the $f$-atom \cite{Bloch_Z2_gauge,Bloch_Z2_gauge_theory}
\begin{equation}
    \hat{H}_{\text{dw}}=-2g\hat{\tau}^z(\hat{a}^\dag_1\hat{a}_2+\hat{a}_2^{\dagger}\hat{a}_1)-2h\hat{\tau}^x.  \label{dw}
\end{equation}
This model exhibits a $\mathbb{Z}_2$ symmetry: the Hamiltonian is invariant under $\hat{\tau}^z \rightarrow -\hat{\tau}^z$ and simultaneously either $\hat{a}_1\rightarrow -\hat{a}_1$ or $\hat{a}_2\rightarrow -\hat{a}_2$.

However, there is a difficulty in extending this model from zero spatial dimension to higher dimensions. If we allow $f$-atoms to hop outside the double wells and their numbers to change, the definition of the $\hat{\bm{\tau}}_{i}$ operator, Eqs.~(\ref{tau_x})--(\ref{tau_z}), becomes invalid, and consequently the $\mathbb{Z}_2$ symmetries are lost. If we allow $a$-atoms to hop outside of the double wells, it will render the local $\mathbb{Z}_2$ symmetries into a global one.

At this stage, two important questions arise for studying ``dynamical gauge fields" with cold atoms. The first one is how to realize dynamical gauge field models with local gauge symmetries, especially, how to extend the double-well models realized in Ref. \cite{Bloch_Z2_gauge} to include the spatial degrees of freedom. The second one is, provided that such dynamical gauge fields models have been realized using cold atoms, what are the unique physical effects, aside from those have been discussed in the context of high energy physics. Recently, various attempts have been made to address these two issues \cite{dynamic_gauge1,dynamic_gauge2,dynamic_gauge3,dynamic_gauge4,dynamic_gauge5,dynamic_gauge6,dynamic_gauge7,dynamic_gauge7_1,dynamic_gauge8,dynamic_gauge9,dynamic_gauge10,dynamic_gauge11,dynamic_gauge12,MBL_dynamic1,MBL_dynamic2,MBL_dynamic3,MBL_dynamic4}.
\begin{figure}[t]
    \centering
    \includegraphics[width=0.8\linewidth]{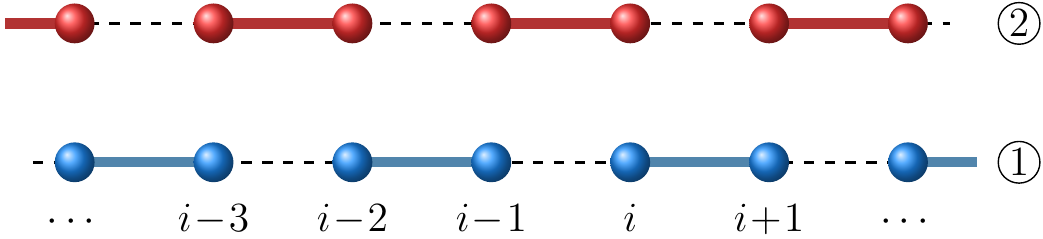}
    \caption{(color online). Our model contains two chains with staggered double wells denoted by the solid blue and red bonds. The blue chain consists of $a$- and $f$-atoms, and the red chain is loaded with $b$- and $d$-atoms. Though for illustration purposes two chains are drawn separated in space, they, in reality, should overlap spatially to allow for interaction between atoms at the same site.}
    \label{scheme}
\end{figure}

In this work we offer our answers to these two questions. Firstly, we propose a scheme to extend the double-well model, Eq.~(\ref{dw}), to one-dimensional. The central idea of our proposal is to utilize the interactions between particles to build up the spatial dimension in a way that the local gauge symmetries are preserved. Secondly, we show that the phenomenon of many-body localization (MBL) can emerge in dynamical gauge field models. Though MBL is usually studied in interacting disordered models \cite{MBL1, MBL2}, we show that the phenomena can also occur in a disorder-free dynamical gauge field model. The key point is that different conserved local gauge charges associated with their local gauge symmetries play the role of disordered potentials in interacting disordered models of MBL, due to the couplings between the gauge fields and the matter fields.

\section{The Model}
\subsection{One-Dimension Dynamical Gauge Fields}
Here we consider two spatially overlapped one-dimensional chains, each of which consists of a series of double wells, as sketched in Fig.~\ref{scheme}. In the first chain, each double well is formed by the $i$-th site and $(i+1)$-th site for \textit{even} number $i$, and each double well contains exactly one $a$-atom and one $f$-atom. This means for each double well with a pair of sites $i$ and $i+1$, we can introduce a spin-$1/2$ operator $\hat {\bm{\tau}}_{i}$ for the $f$-atom
\begin{align}
    \hat{\tau}_{i}^x=\frac{1}{2}\left(\hat{f}^\dag_i\hat{f}_{i+1}+\hat{f}^\dag_{i+1}\hat{f}_i\right), \label{tau_x1}\\
    \hat{\tau}_{i}^y=\frac{i}{2}\left(\hat{f}^\dag_{i+1}\hat{f}_i-\hat{f}^\dag_i\hat{f}_{i+1}\right), \label{tau_y1} \\
    \hat{\tau}_{i}^z=\frac{1}{2}\left(\hat{f}^\dag_i\hat{f}_{i}-\hat{f}^\dag_{i+1}\hat{f}_{i+1}\right).\label{tau_z1}
\end{align}
Similarly, we can introduce another spin-$1/2$ operator $\hat{\bm{\sigma}}_{i}$ for the $a$-atom
\begin{align}
    \hat{\sigma}_{i}^x=\frac{1}{2}\left(\hat{a}^\dag_i\hat{a}_{i+1}+\hat{a}^\dag_{i+1}\hat{a}_i\right), \label{sigma_x1}\\
    \hat{\sigma}_{i}^y=\frac{i}{2}\left(\hat{a}^\dag_{i+1}\hat{a}_i-\hat{a}^\dag_i\hat{a}_{i+1}\right), \label{sigma_y1} \\
    \hat{\sigma}_{i}^z=\frac{1}{2}\left(\hat{a}^\dag_i\hat{a}_{i}-\hat{a}^\dag_{i+1}\hat{a}_{i+1}\right).\label{sigma_z1}
\end{align}
In this way, we have introduced $\hat{\bm{\tau}}_{i}$ and $\hat{\bm{\sigma}}_{i}$ for every even number $i$. The Hamiltonian of this chain is simply a sum of that of each individual double well, Eq.~(\ref{dw}), and takes the following form
\begin{equation}
    \hat{H} = \sum\limits_{i=\text{even}}-4g\hat{\tau}_{i}^z\hat{\sigma}_{i}^x-2h\hat{\tau}_{i}^x, \label{H_chain}
\end{equation}
in terms of previously defined spin-$1/2$ operators. This Hamiltonian has included the effect of lattice shaking, the hopping of both $a$- and $f$-atoms inside the double wells, and the interactions between $a$- and $f$-atoms \cite{Bloch_Z2_gauge,Bloch_Z2_gauge_theory}. It is exactly the same model as that has been realized in the recent experiment \cite{Bloch_Z2_gauge}.

Next, we consider the second chain constituted with $b$- and $d$-atoms that play the same role as $a$- and $f$-atoms in the first chain respectively. In this chain, each double well is formed by the $i$-th and $(i+1)$-th sites with \textit{odd} $i$. Similar to the first chain, these double wells are disconnected and each double well contains exactly one $b$-atom and one $d$-atom. We can, therefore, introduce two spin-$1/2$ operators $\hat{\bm{\tau}}_i$ and $\hat{\bm{\sigma}}_i$ for every odd number $i$. Their definitions are given by Eqs.~(\ref{tau_x1})--(\ref{tau_z1}) 
upon substituting $f$-atom operators with $d$-atom operators and by Eqs.~(\ref{sigma_x1})--(\ref{sigma_z1}) after replacing $a$-atom operators with $b$-atom operators. In this way, we have introduced $\hat{\bm{\tau}}_{i}$ and $\hat{\bm{\sigma}}_{i}$ for every odd number $i$. Moreover, the same shaking protocol is applied to this chain, which generates the same Hamiltonian as Eq.~(\ref{H_chain}) except that the summation is over odd number of $i$. At this point, we have defined two sets of spin operators $\hat{\bm{\sigma}}_{i}$ and $\hat{\bm{\tau}}_i$ for all links in this one-dimensional system.

Now we turn on interactions between the two chains. We emphasize again that these two chains are arranged to be spatially overlapping with each other such that atoms at the same site can interact with each other. The interactions between $a$- and $f$-atoms and between $b$- and $d$-atoms have already been taken into account when writing down the effective Hamiltonian Eq.~(\ref{H_chain}). Among the rest interactions, we consider the case where the dominate contribution comes from the interactions between $a$- and $b$-atoms, $U\sum_{i}\hat{a}^\dag_i\hat{a}_i\hat{b}^\dag_i\hat{b}_i$.
Noting that for even number $i$, we have
\begin{align}
	\hat{a}^\dag_i\hat{a}_i & = 1/2 + \hat{\sigma}_{i}^z;  \\
	\hat{b}^\dag_i\hat{b}_i & = 1/2 - \hat{\sigma}_{i-1}^z.
\end{align}
And for odd number $i$, we have
\begin{align}
	\hat{a}^\dag_i\hat{a}_i &= 1/2 - \hat{\sigma}_{i-1}^z; \\
	\hat{b}^\dag_i\hat{b}_i &= 1/2 + \hat{\sigma}_{i}^z.
\end{align}
The interactions between $a$- and $b$-atoms can then be cast into the following form
\begin{equation}
    U\sum_{i}\hat{a}^\dag_i\hat{a}_i\hat{b}^\dag_i\hat{b}_i = - U\sum\limits_{i} (\hat{\sigma}^{z}_{i}\hat{\sigma}^z_{i+1} - 1/4) \, .
\end{equation}
Hence, the total Hamiltonian of this system reads
\begin{equation}
    \hat{H}=\sum\limits_{i} \left( -4g\hat{\tau}_{i}^z\hat{\sigma}_{i}^x-2h\hat{\tau}_{i}^x-4\hat{\sigma}_{i}^z\hat{\sigma}_{i+1}^z \right) . \label{H_sys}
\end{equation}
Here and hereafter, we set $U/4=1$ as our energy unit, and $g$ and $h$ here are actually dimensionless numbers $4g/U$ and $4h/U$ in terms of the original parameters.
Obviously, this Hamiltonian is invariant under the transformation $\hat{\sigma}_{i}^x \rightarrow - \hat{\sigma}_{i}^x$ and $\hat{\tau}_{i}^z \rightarrow - \hat{\tau}_{i}^z$, and possesses a local $\mathbb{Z}_{2}$ gauge symmetry.
\begin{figure}[t]
    \centering
    \includegraphics[width=0.8\linewidth]{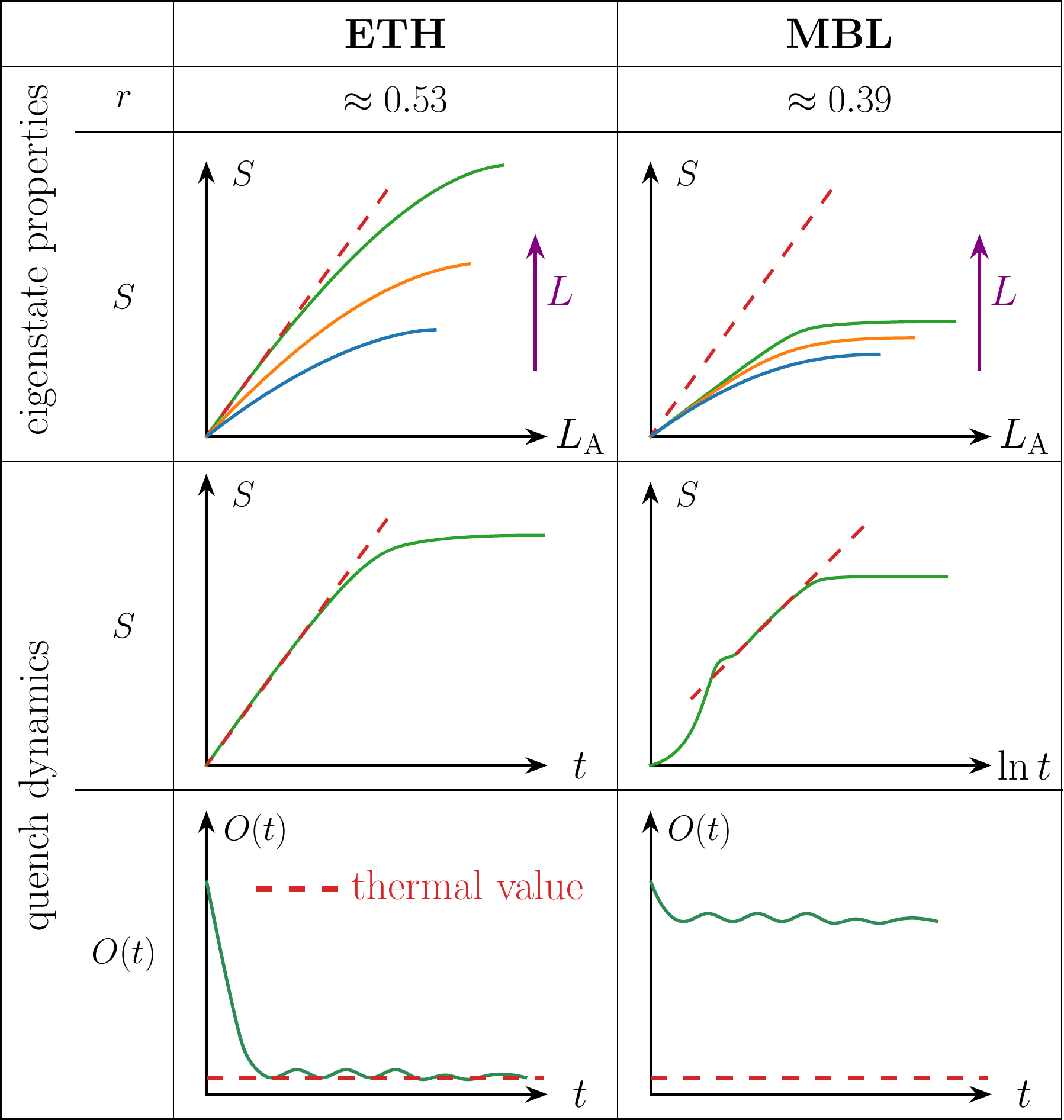}
    \caption{(color online). Summary of four metrics to distinguish a thermal (ETH) phase from a MBL phase: i) $r$,  which is a quantity characterizing level statistics of eigenstates and is defined by Eq.~(\ref{r_value}); ii) subsystem size $L_{\text{A}}$ dependence of the entanglement entropy $S$ of an eigenstate with non-zero energy density; iii) time evolution of the entanglement entropy of subsystem $A$ after a quench from a product state of different sites in linear time scale (left panel) and logarithmic time scale (right panel); iv) time evolution of a physical observable $\hat{O}$ after a quench from a generic initial state.}
    \label{ETH_MBL}
\end{figure}

\subsection{Mapping to a Model of Many-Body Localization}
For later convenience, we shall rotate $\hat{\bm{\tau}}_i$ as follows:
\begin{equation} \label{eq:tau_rotation}
    \hat{\tau}^x_i\rightarrow \hat{\tau}^z_i, \   \ \hat{\tau}^y_i\rightarrow -\hat{\tau}^y_i, \text{ and } \hat{\tau}^z_i\rightarrow \hat{\tau}^x_i \, .
\end{equation}
The Hamiltonian then becomes
\begin{equation}
    \hat{H}=\sum\limits_{i}\left(-4g\hat{\tau}_{i}^x\hat{\sigma}_{i}^x-2h\hat{\tau}_{i}^z-4\hat{\sigma}_{i}^z\hat{\sigma}_{i+1}^z \right) \, . \label{H_sys}
\end{equation}
In this new basis, one can easily see that $\hat{Q}_i = 4\hat{\sigma}_{i}^z \hat{\tau}_{i}^z$ commutes with $\hat{H}$ and its eigenvalue $q_{i}$ takes the value of $+1$ or $-1$. For any eigenstate $|\rangle$, we have $\hat{Q}_i |\rangle = q_i |\rangle$ and $q_{i}$ can be understood as the conserved local gauge charge associated with the local gauge symmetries. 
The four local bases $|\sigma_i^z \tau_i^z\rangle$ of site $i$ fall into two classes: $|\uparrow\uparrow\rangle$ and $|\downarrow\downarrow\rangle$ with $q_{i}=1$, and $|\uparrow\downarrow\rangle$ and $|\downarrow\uparrow\rangle$ with $q_{i}=-1$.
To make use of the reduced local Hilbert space for given $q_{i}$, we introduce another spin-$1/2$ operator $\hat{\bm{\Gamma}}_i$ defined as
\begin{equation}
    \hat{\Gamma}_i^x= 2 \hat{\tau}^x_i\hat{\sigma}_i^x, \quad \hat{\Gamma}_i^y= 2 \hat{\tau}^x_i\hat{\sigma}^y_i, \quad \text{and} \quad \hat{\Gamma}_i^z=\hat{\sigma}_i^z.
\end{equation}
Indeed, these operators satisfy the $SU(2)$ algebra: $[\hat{\Gamma}_i^x, \hat{\Gamma}_i^y] = i \hat{\Gamma}_i^z$, $[\hat{\Gamma}_i^y, \hat{\Gamma}_i^z]=i\hat{\Gamma}_i^x$ and $[\hat{\Gamma}_i^z, \hat{\Gamma}_i^x]=i\hat{\Gamma}_i^y$ (note $\hbar = 1$). 
It can be verified that the spin-up and spin-down states, denoted as $|\Uparrow\rangle$ and $|\Downarrow\rangle$, of the $\hat{\bm{\Gamma}}_{i}$ operator, are $|\uparrow\uparrow\rangle$ and $|\downarrow\downarrow\rangle$ respectively for subspace with $q_i=1$, and are $|\uparrow\downarrow\rangle$ and $|\downarrow\uparrow\rangle$ respectively for subspace with $q_{i}=-1$. 
Noticing that
\begin{equation}
    \hat{\tau}^z_i = 4\hat{\tau}^z_i\hat{\sigma}^z_i\hat{\sigma}^z_i= \hat{Q}_i\hat{\Gamma}^z_i \, ,
\end{equation}
we can always replace $\hat{\tau}_i$ by $q_i \hat{\Gamma}_{i}$ when acting the Hamiltonian on eigenstates.
With the help of the definition of the $\hat{\bm{\Gamma}}_{i}$ operators, the Hamiltonian can be simplified as 
\begin{equation}
    \hat{H}=\sum\limits_{i}\left( -2g\hat{\Gamma}_{i}^x-2h q_i\hat{\Gamma}_{i}^z-4\hat{\Gamma}_{i}^z\hat{\Gamma}_{i+1}^z \right). \label{H_MBL}
\end{equation}
Because $q_i$ takes $\pm 1$ in different symmetry sectors,  $hq_i$ can be either $+h$ or $-h$ and acts as a $\mathbb{Z}_2$ random field. In this way, we map our model to a transverse Ising model in the presence of a disordered longitudinal field. This mapping clearly reveals that the local conserved gauge charges play the role of disordered potentials. Since the gauge charge is always quantized, the difference between this model and usual models of MBL is the distribution of the random field. In the usual models of MBL, the field takes a continuous distribution, say, uniformly distributed in $[-h, h]$, whereas in our effective model, the field takes two discrete values with equal probabilities.
\begin{figure}[t]
    \centering
    \includegraphics[width=0.8\linewidth]{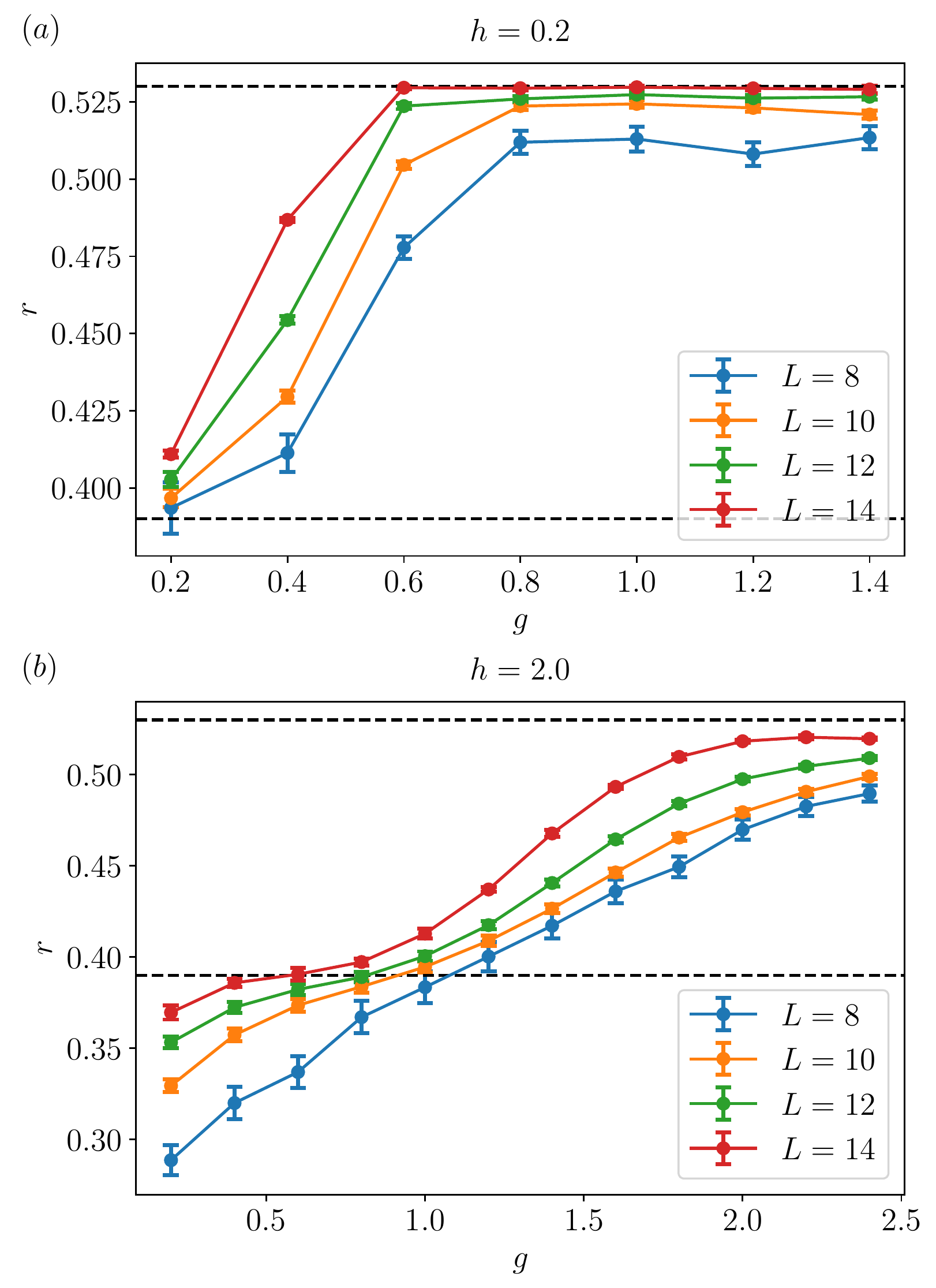}
    \caption{(color online). The value of $r$, defined in Eq.~(\ref{r_value}), as a function of $g$ for system sizes
    $L=8, 10, 12$ and $14$ at (a) $h=0.2$ and (b) $h=2.0$. The error bar stands for one standard deviation when averaging over symmetry sectors. The two dashed black lines are $r=0.39$ and $r=0.53$ lines.}
    \label{r_value_plot}
\end{figure}
Nevertheless, this difference does not exclude the possibility of MBL. In the weak disorder limit $h \ll g$, the system is ergodic and obeys the eigenstate thermalization hypothesis (ETH) \cite{ETH1, ETH2, ETH3,Rigol:AIP}. In the opposite strong disorder limit $h \gg g$, ETH fails and the system is expected to be in the MBL regime \cite{MBL1, MBL2}.
Below we will demonstrate the presence of both the ETH (thermal) regime and the MBL regime in this model from several different metrics.

\section{Many-Body Localization}
The differences between the ETH regime and the MBL regime can be characterized using a number of different physical quantities \cite{metrics,Huse:ARCMP,Rigol:AIP,Abanin:RMP}. In this section, we will employ four different metrics, related to properties of eigenstate and behavior of quench dynamics from a product state listed in Fig.~\ref{ETH_MBL}.
\begin{figure}[t]
    \centering
    \includegraphics[width=0.8\linewidth]{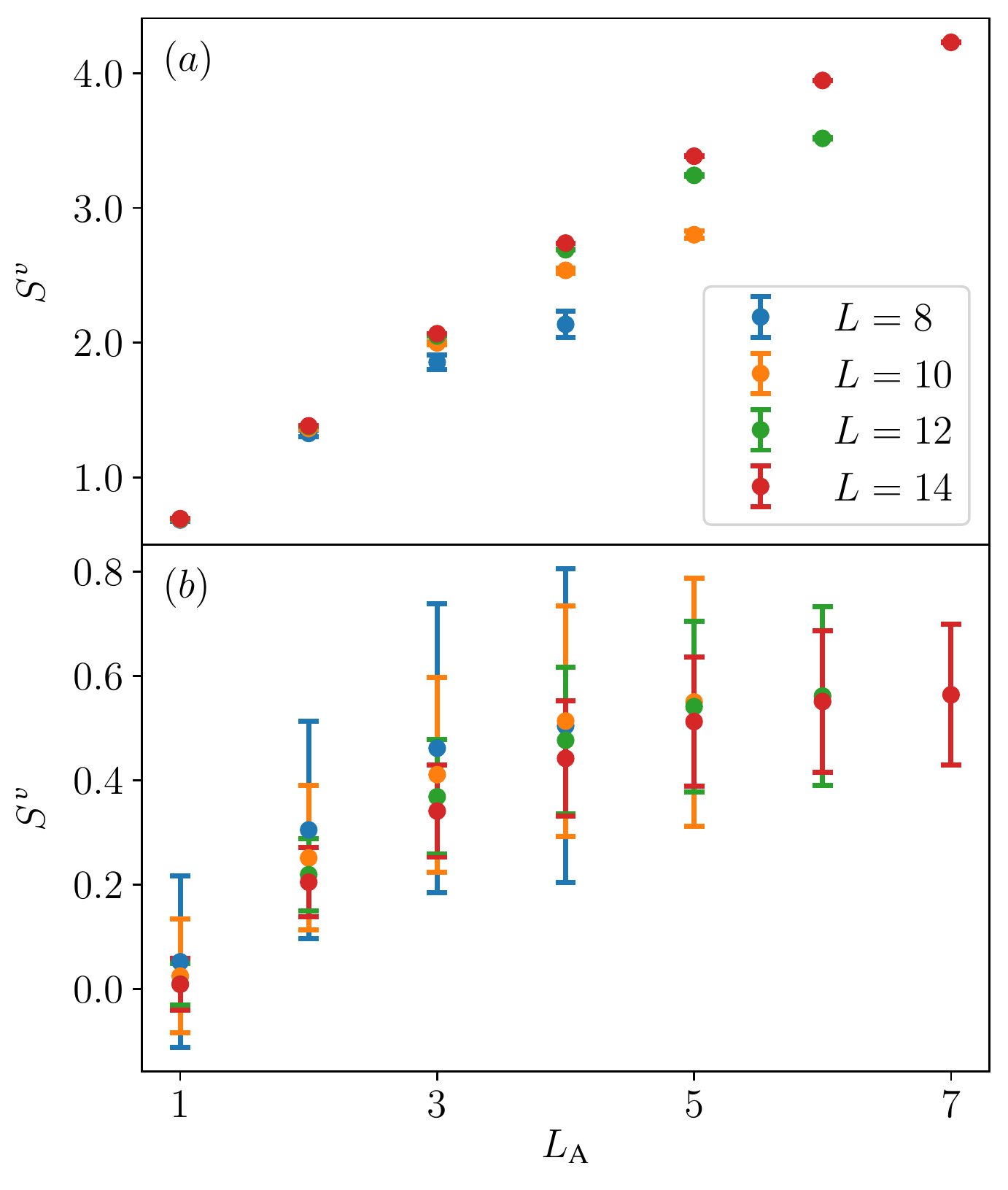}
    \caption{(color online). The von Neumann entanglement entropy $S^{v}$ as a function of subsystem size $L_{\text{A}}$ averaged over eigenstates within a small energy window $[E, E + \Delta E]$, where $E=-1$ and $\Delta E = 0.1$, at (a) $(g, h)=(1.0, 0.2)$ and (b) $(g, h)=(0.2, 2.0)$. Data for system sizes $L=8, 10, 12$ and $14$ are presented, and error bars stand for one standard deviation.}
    \label{eigen_entropy}
\end{figure}

\subsection{Level Statistics}
We first focus on the eigenstate properties. It is known that in the ETH regime the energy level spacing obeys the Wigner--Dyson distribution and in the MBL regime the energy level spacing obeys the Poisson distribution \cite{Rigol:AIP}.
To determine the specific distribution, a dimensionless quantity $r$ has been introduced \cite{r}
\begin{equation}
    r = \langle r_{n}\rangle= \left\langle\frac{\min(\delta_{n}, \delta_{n-1})}{\max(\delta_{n}, \delta_{n-1})}\right\rangle , \label{r_value}
\end{equation}
where $n$ is the ascending energy level index, $\delta_{n} = E_{n+1}-E_{n}$ is the energy level spacing between two consecutive energy levels, and the average, in our case, is taken over all energy levels, including a first average over all eigenstates in a given symmetry sector and then a second average over different symmetry sectors. It can be shown that $r \approx 0.53$ for the Wigner--Dyson distribution and $r \approx 0.39$ for the Poisson distribution \cite{r}.

We use the exact diagonalization method with the open boundary condition to compute all eigenvalues and determine the value of $r$. As can been seen from Fig.~\ref{r_value_plot}, the general trend is that when $g$ becomes larger, the value of $r$ approaches $0.53$, and when $g$ is small, the value of $r$ is around $0.39$. By comparing Fig.~\ref{r_value_plot} (a) and (b), one can see that when $h$ is larger, it also requires a larger $g$ to reach the ETH regime where $r\approx 0.53$. This agrees with our expectation that the system is driven from the ETH regime to the MBL regime as the relative strength of disorder increases.
\begin{figure}[t]
    \centering
    \includegraphics[width=0.8\linewidth]{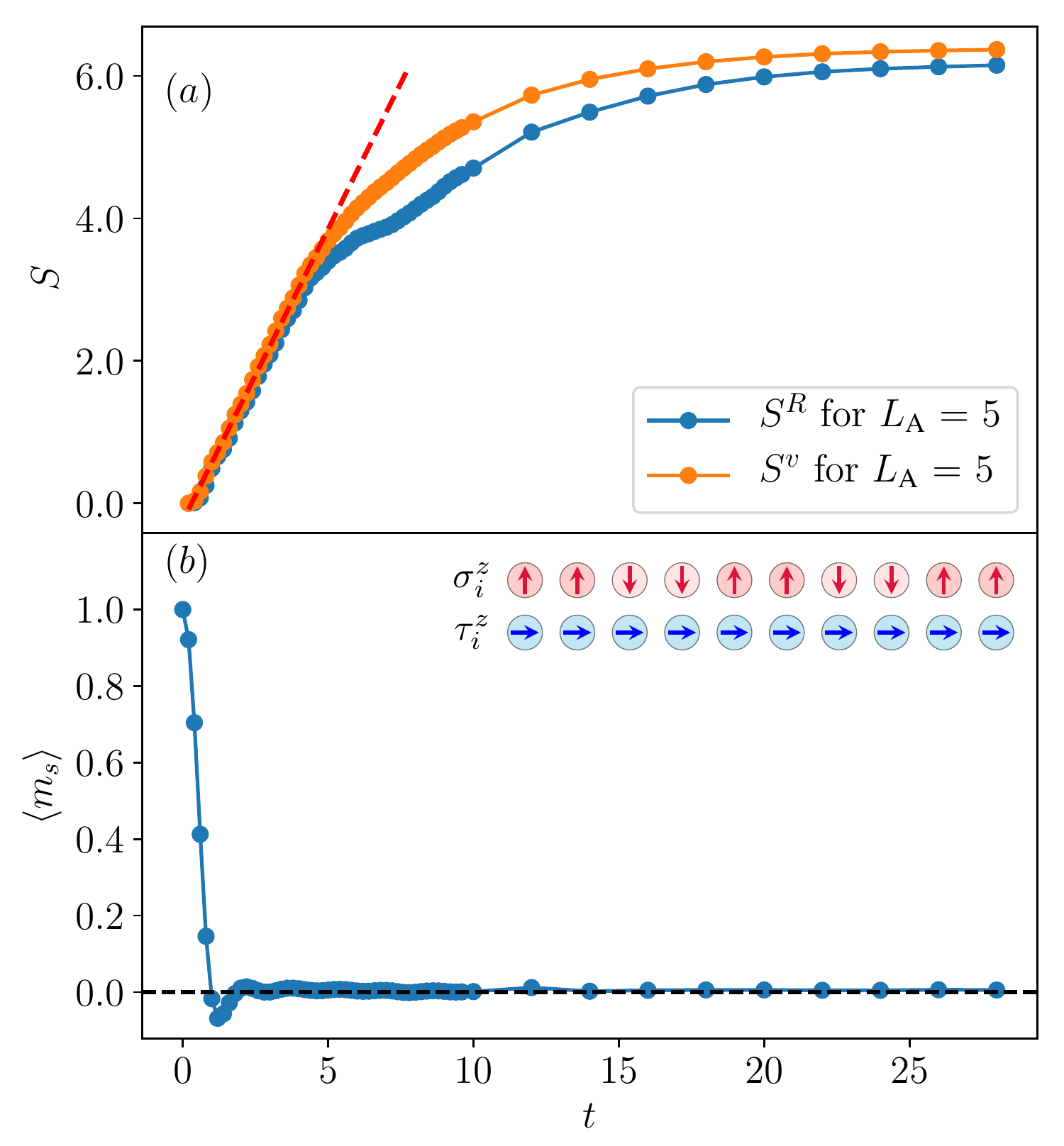}
    \caption{(color online). Quench dynamics of (a) entanglement entropies and (b) the staggered magnetization defined in Eq.~(\ref{ms}) from an initial product state illustrated in the inset of (b). Both the von Neumann entropy $S^v$ and the R\'enyi entropy $S^R$ are plotted. Here $t$ actually stands for dimensionless $4tU$ in terms of original parameters. The system size is $L=10$ and the size of subsystem $A$ is $L_{A}=5$. The simulation is performed at $(g, h)=(1.0, 0.2)$ which is shown to be inside the ETH regime.}
    \label{quench_ETH}
\end{figure}

\subsection{Entanglement Entropies for Eigenstates}

How the von Neumann entanglement entropy $S^{v}$ of a generic eigenstate with non-zero energy density scales with the size of the subsystem, $L_{\text{A}}$, is another way to differentiate the ETH regime from the MBL regime. The ETH ansatz implies the reduced density matrix of a small subsystem $A$ approaches its thermal density matrix at the temperature $T$ fixed by the energy of the eigenstate, leading to the volume law for $S^{v}$ \cite{Abanin:RMP,Huse:ARCMP}.
In contrast, a MBL system fails to thermalize, and the entanglement entropy obeys the area law \cite{Huse:ARCMP,Abanin:RMP}. 

The above difference is schematically shown in Fig.~\ref{ETH_MBL}, and the behavior of $S^{v}$ is also verified in our numerical studies. 
As can be seen from Fig.~\ref{eigen_entropy}(a), when $h$ is small compared with $g$, the entanglement entropy increases almost linearly with $L_{\text{A}}$ until $L_{\text{A}}$ approaches half of the system size, confirming the volume law. In contrast, as shown in Fig.~\ref{eigen_entropy}(b), when $h$ is large compared with $g$, the entanglement entropy tends to quickly saturates as $L_{\text{A}}$ increases, demonstrating the area law.
\begin{figure}[t]
    \centering
    \includegraphics[width=0.8\linewidth]{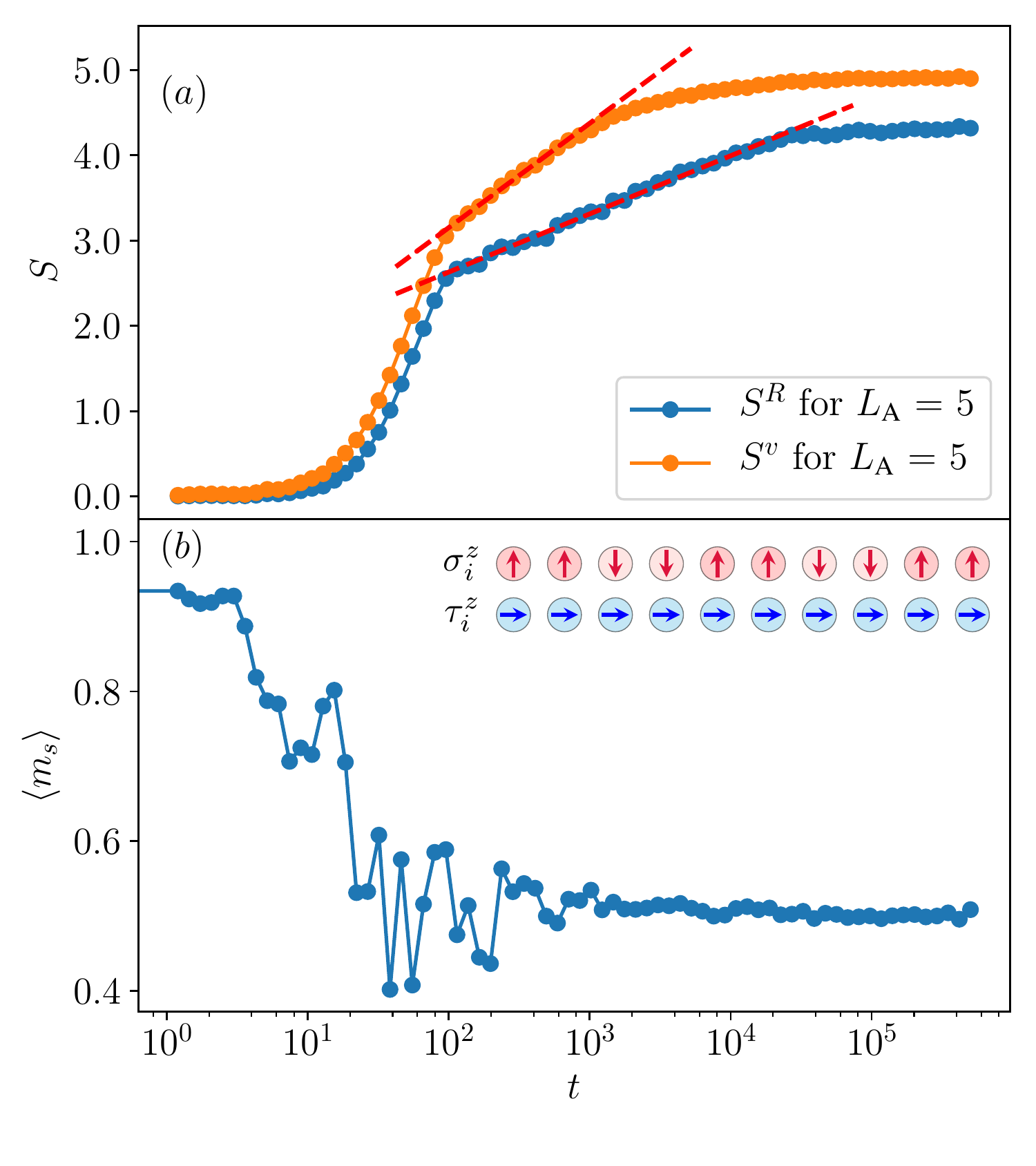}
    \caption{(color online). Quench dynamics of (a) entanglement entropies and (b) the staggered magnetization defined in Eq.~(\ref{ms}) from an initial product state illustrated in the inset of (b). Both the von Neumann entropy $S^v$ and the R\'enyi entropy $S^R$ are plotted in logarithmic time scale. Here $t$ actually stands for dimensionless $4tU$ in terms of original parameters. The system size is $L=10$ and the size of subsystem $A$ is $L_{A}=5$. The simulation is performed at $(g, h)=(0.2, 1.0)$ which is shown to be inside the MBL regime.
    }
    \label{quench_MBL}
\end{figure}

\subsection{Entanglement Entropy Growth after a Quench}
Aside from eigenstate properties, the differences between the ETH regime and the MBL regime also manifest themselves in quench dynamics. Starting with a product state of different sites, the entanglement entropy for subsystem $A$ will increase from zero. In the ETH regime, the entanglement entropy increases linearly until it saturates  \cite{linear_growth}. In the MBL regime, following a short linear growth, the entanglement entropy will undergo a slow logarithmically increase in an extended time region \cite{log_growth1,log_growth2,log_growth3,log_growth4}. This logarithmic increase is often seen as the hallmark of the MBL phase and can be explained by the phenomenological theory of MBL \cite{phenomenology_MBL1,phenomenology_MBL2}. Yet this linear versus logarithmic growth of entropy is also directly related to the exponential versus power-law behavior of the out-of-time-ordered correlation \cite{Zhai_OTOC,OTOC1,OTOC2,OTOC3,OTOC4,OTOC5}. 

Our initial product state is prepared with all $\bm{\tau}_{i}$ spins polarized along the $x$-direction and all $\bm{\sigma}_{i}$ spins aligned upwards or downwards in a domain-wall fashion, as illustrated in the inset of Fig.~\ref{quench_ETH}(b) and Fig.~\ref{quench_MBL}(b). In terms of the original $\bm{\tau}_{i}$ spins before rotation, Eq.~(\ref{eq:tau_rotation}), all $\bm{\tau}_{i}$ spins are actually polarized along the $z$-direction. That is to say all the $f$- and $d$-atoms are prepared in the left wells, and all the $a$- and $b$-atoms are localized either in the left ($\sigma_{i}^{z}$ upwards) or right wells ($\sigma_{i}^{z}$ downwards). The advantage of choosing such an initial state is twofold. Firstly, the wave function is an equal weight superposition of states from all symmetry sectors. Secondly, the initial state has a relative high energy which mitigates the problem of limited accessible system sizes in numerical simulations.

We numerically study the growth of both the von Neumann entanglement entropy $S^v$ and the second-order R\'{e}nyi entropy $S^R$ of the subsystem $A$ that is chosen as half of the entire system. The presence of both the ETH regime and the MBL regime is demonstrated using two representative parameter sets, $(g, h) =(1.0, 0.2)$ and $(g, h) =(0.2, 1.0)$, as shown in Fig.~\ref{quench_ETH} and Fig.~\ref{quench_MBL} respectively. From Fig.~\ref{quench_ETH}(a), one can see that for the former case where the disorder ($h$-term) is relatively weak, two entanglement entropies both increase linearly in time, as indicated by the red dashed line. 
In contrast, for the latter case where the disorder is relatively strong, a logarithmic growth of both entanglement entropies is revealed, as demonstrated by the red dashed lines in Fig.~\ref{quench_MBL}(a). The characteristic saturation time is also much longer than the former case.

\subsection{Evolution of Observable after a Quench}
Finally, we study the time evolution of physical observables following a quench. If a system is thermal, all local physical observables will evolve toward their thermal equilibrium values. For a MBL system, local physical observables, however, keep part of the memory of the initial state and do not necessarily approach the thermal equilibrium values. This criterion has been applied in identifying MBL phases in cold atom experiments \cite{Bloch_MBL}.

Here we consider a special observable, the staggered magnetization, whose definition depends on the initial state, 
\begin{equation}
    \hat{m}_{s} = \frac{2}{L} \sum_{i} \xi_{i} \hat{\sigma}_{i}^{z}, \label{ms}
\end{equation}
where $\xi_{i}=1$ if initially $\sigma_{i}^{z}=1/2$ and $\xi_i=-1$ if initially $\sigma_{i}^{z}=-1/2$.
The advantage of defining such an observable is also twofold. First, initially, the expectation value of our observable is always $m_{s}=1$. Secondly, it is easy to show that the thermal average of $\hat{m}_{s}$ is always zero because of the $\hat{\sigma}_{i}^{z} \rightarrow -\hat{\sigma}_{i}^{z}, q_{i} \rightarrow -q_{i}$ symmetry of the Hamiltonian, under which $m_s$ changes its sign. By comparing Fig.~\ref{quench_ETH}(b) and Fig.~\ref{quench_MBL}(b), it is clear that the system thermalizes in the former case but fails to thermalize in the latter case.

\section{Outlook}
In summary, we have established that many-body localization can occur in a disorder-free system with local gauge symmetries. In our model, we explicitly show that gauge charges of different symmetry sectors can play the role of disorder potentials. Recently, several works have also studied MBL in different dynamical gauge theories with metrics different  from our work \cite{MBL_dynamic1,MBL_dynamic2,MBL_dynamic3,MBL_dynamic4}. All together with these efforts, we would like to conclude this work by posting the following question. Whether the many-body localization is a generic feature of all dynamical gauge theories with local gauge symmetries? If yes, whether this new development can also shed light on studying lattice gauge theories in high energy physics.

\textit{Acknowledgment.} This work is supported by Beijing Outstanding Young Scientist Program, MOST under Grant No. 2016YFA0301600 and NSFC Grant No. 11734010.

\end{document}